\documentclass{ws-procs9x6}

\begin{document}

\title{Parameter degeneracy and reactor experiments
\footnote{\uppercase{T}his work is supported in part by
the \uppercase{G}rant-in-\uppercase{A}id for \uppercase{S}cientific
\uppercase{R}esearch
in \uppercase{P}riority \uppercase{A}reas \uppercase{N}o. 12047222
and \uppercase{N}o. 13640295,
\uppercase{J}apan \uppercase{M}inistry
of \uppercase{E}ducation, \uppercase{C}ulture, \uppercase{S}ports,
\uppercase{S}cience, and \uppercase{T}echnology.}}

\author{Osamu Yasuda}

\address{Department of Physics, Tokyo Metropolitan University,\\
1-1 Minami-Osawa,
Hachioji, Tokyo 192-0397, Japan\\
E-mail: yasuda@phys.metro-u.ac.jp}


\maketitle

\abstracts{
Degeneracies of the neutrino oscillation parameters
are explained using the $\sin^22\theta_{13}$--$s^2_{23}$ plane.
Measurements of $\sin^22\theta_{13}$ by reactor experiments
are free from
the parameter degeneracies which occur in accelerator
appearance experiments, and reactor experiments
play a role complementary to accelerator experiments.
It is shown that the reactor measurement may be able to
resolve the degeneracy in $\theta_{23}$
if $\sin^22\theta_{13}$ and $\cos^22\theta_{23}$
are relatively large.
}

\section{Introduction}
Thanks to the successful experiments on atmospheric and
solar neutrinos and KamLAND, we now know approximately
the values of the mixing angles and the mass squared differences
of the atmospheric and solar neutrino oscillations:
$(\sin^22\theta_{12}, \Delta m^2_{21})\simeq
(0.8, 7\times10^{-5}{\rm eV}^2)$ for the solar neutrino
and $(\sin^22\theta_{23}, |\Delta m^2_{31}|)\simeq
(1.0, 3\times10^{-3}{\rm eV}^2)$ for the atmospheric neutrino.
In the three flavor framework of neutrino oscillations,
the quantities which are still unknown to date are
the third mixing angle $\theta_{13}$, the sign
of the mass squared difference $\Delta m^2_{31}$ 
of the atmospheric neutrino oscillation, and
the CP phase $\delta$.
Among these three quantities, the determination of
$\theta_{13}$ is the next goal in the near future
neutrino experiments.

In this talk I will first explain briefly
the ambiguity due to the parameter degeneracies, which
occur in the long baseline experiments, using the
$\sin^22\theta_{13}$--$s^2_{23}$ plane, and then
I will show that reactor experiments will play
a role complementary to
accelerator experiments, and it may resolve a certain
degeneracy when combined with an accelerator experiment.
This talk is based on the work\cite{Minakata:2002jv},
in which references on the subject can be found.

\begin{figure}
\vglue .1cm
\hglue -.1cm
\includegraphics[scale=0.3]{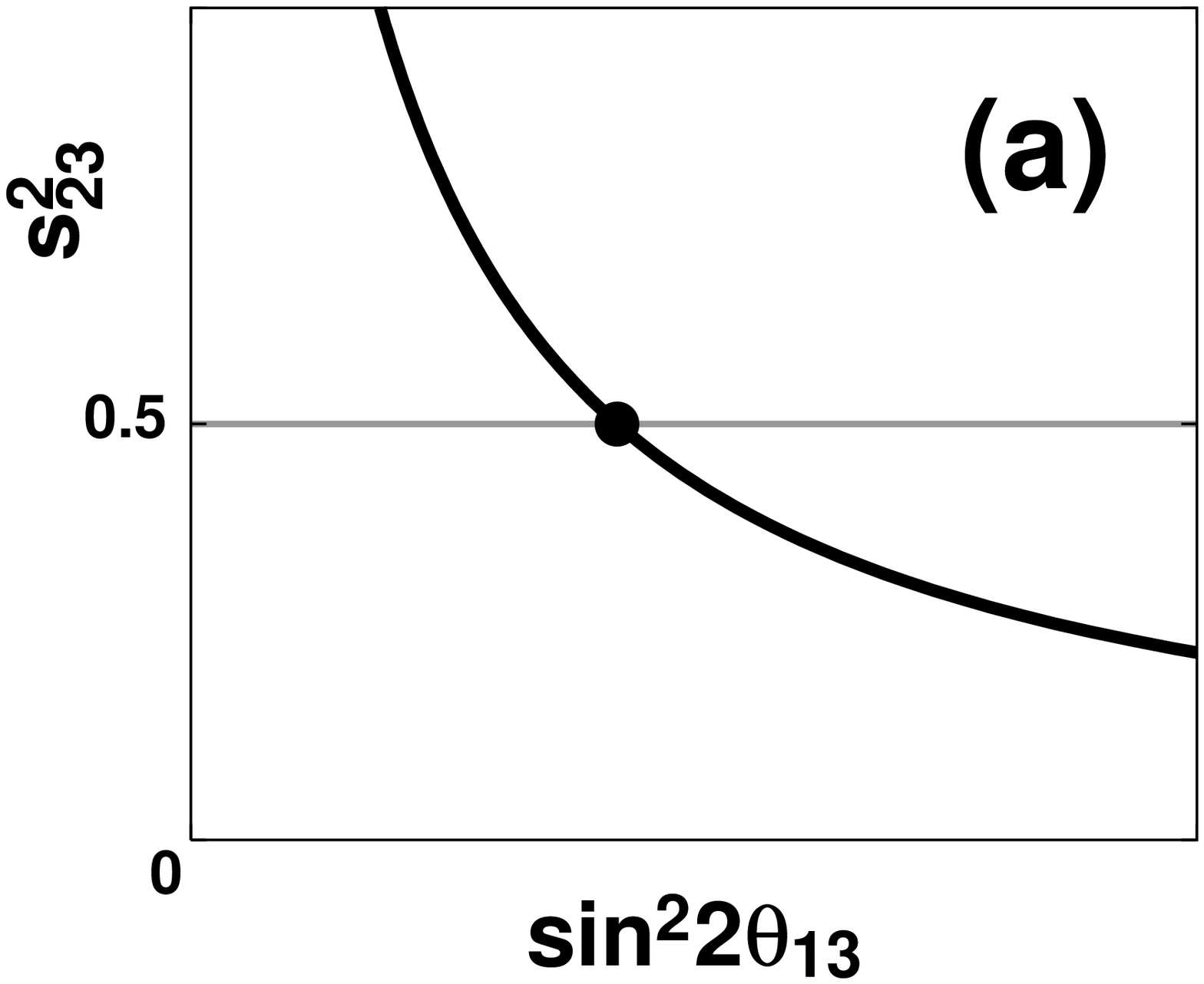}
\vglue -4.3cm
\hglue 5.5cm
\includegraphics[scale=0.3]{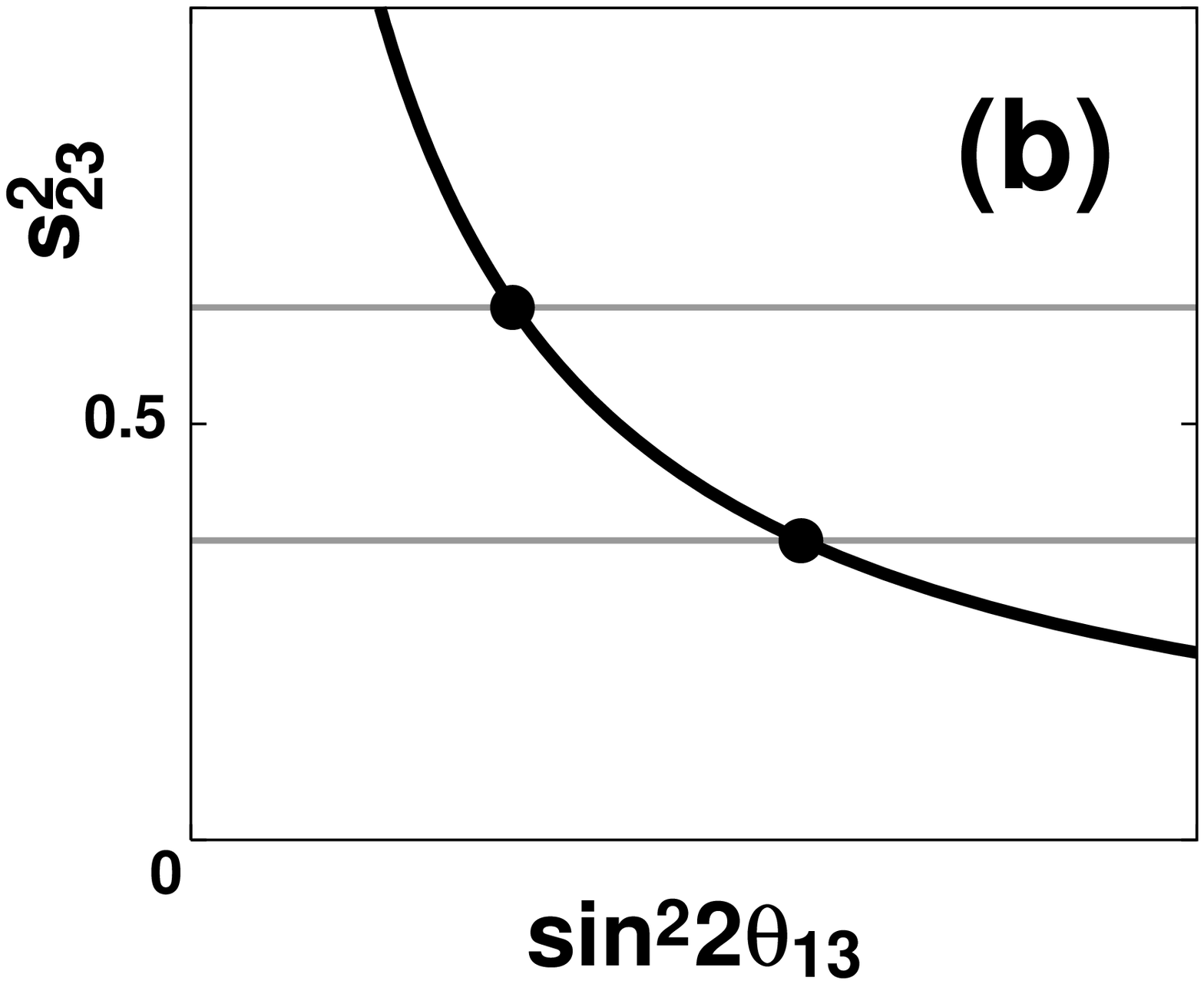}
\vglue .5cm
\hglue -.1cm
\includegraphics[scale=0.3]{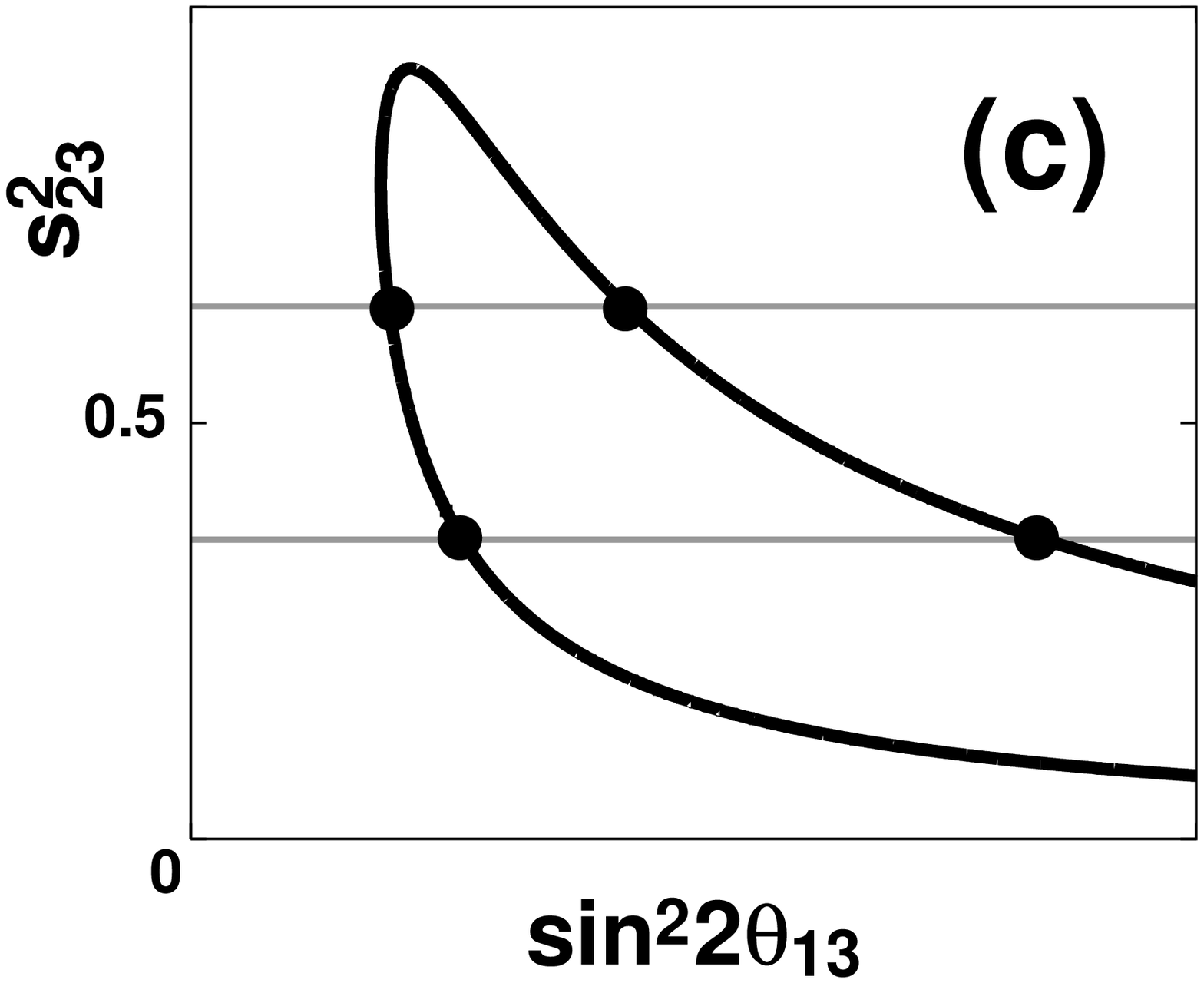}
\vglue -4.3cm
\hglue 5.5cm
\includegraphics[scale=0.3]{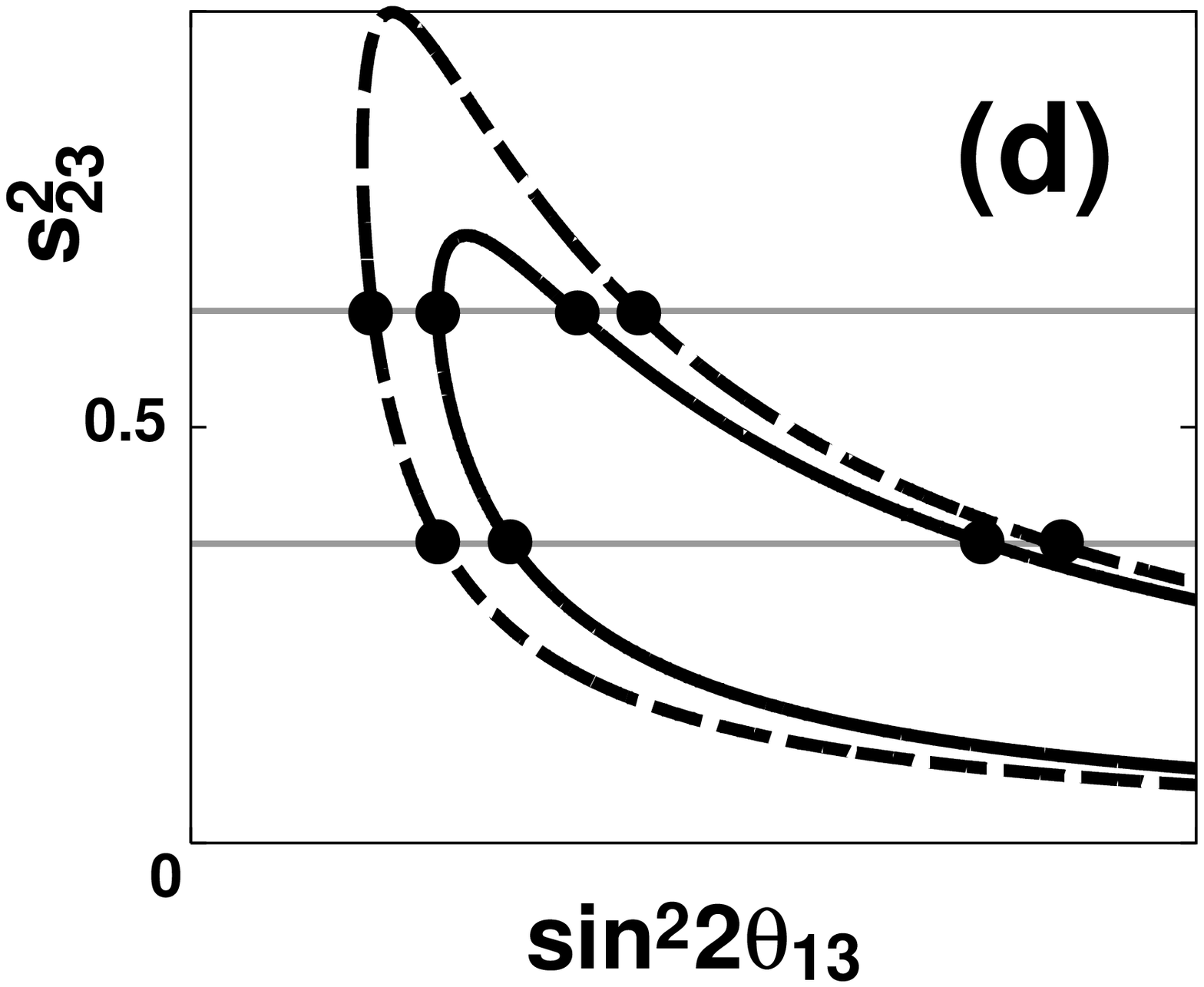}
\vglue .5cm
\hglue -.1cm
\includegraphics[scale=0.3]{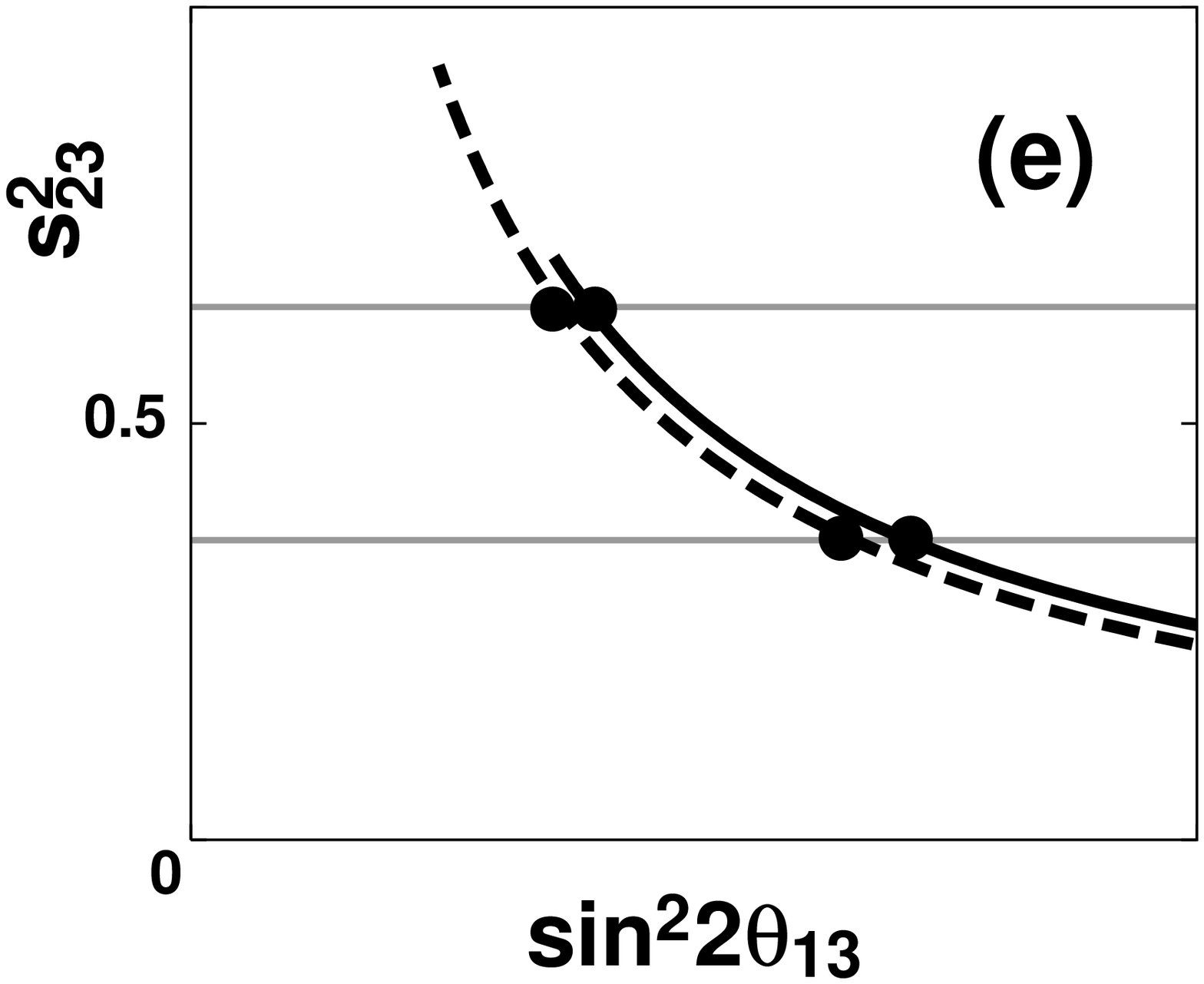}
\vglue 0.5cm
\caption{{\scriptsize Contours which are given
when both $P$ and ${\bar P}$ are known.
The eight-fold parameter degeneracy is lifted
as the small parameters are switched on:
($\cos^22\theta_{23}$, $|\Delta m_{21}^2/\Delta m_{31}^2|$, $AL$)
(a)($=0$, $=0$, $=0$),
(b)($\ne0$, $=0$, $=0$),
(c)($\ne0$, $\ne0$, $=0$),
(d)($\ne0$, $\ne0$, $\ne0$),
(e)($\ne0$, $\ne0$, $\ne0$, at the first oscillation
maximum).
The solid (dashed) line stands for
the case for $\Delta m^2_{31}>0$
($\Delta m^2_{31}<0$), respectively,
(a),(b),(c).
${\bar P}=P ({\bar P}>P)$ is assumed
in (a), (b) and (c) ((d) and (e)), respectively.
}}
\label{fig:degeneracy}
\end{figure}

\begin{figure}
\vglue -2.5cm
\hglue -1.2cm
\includegraphics[scale=0.36]{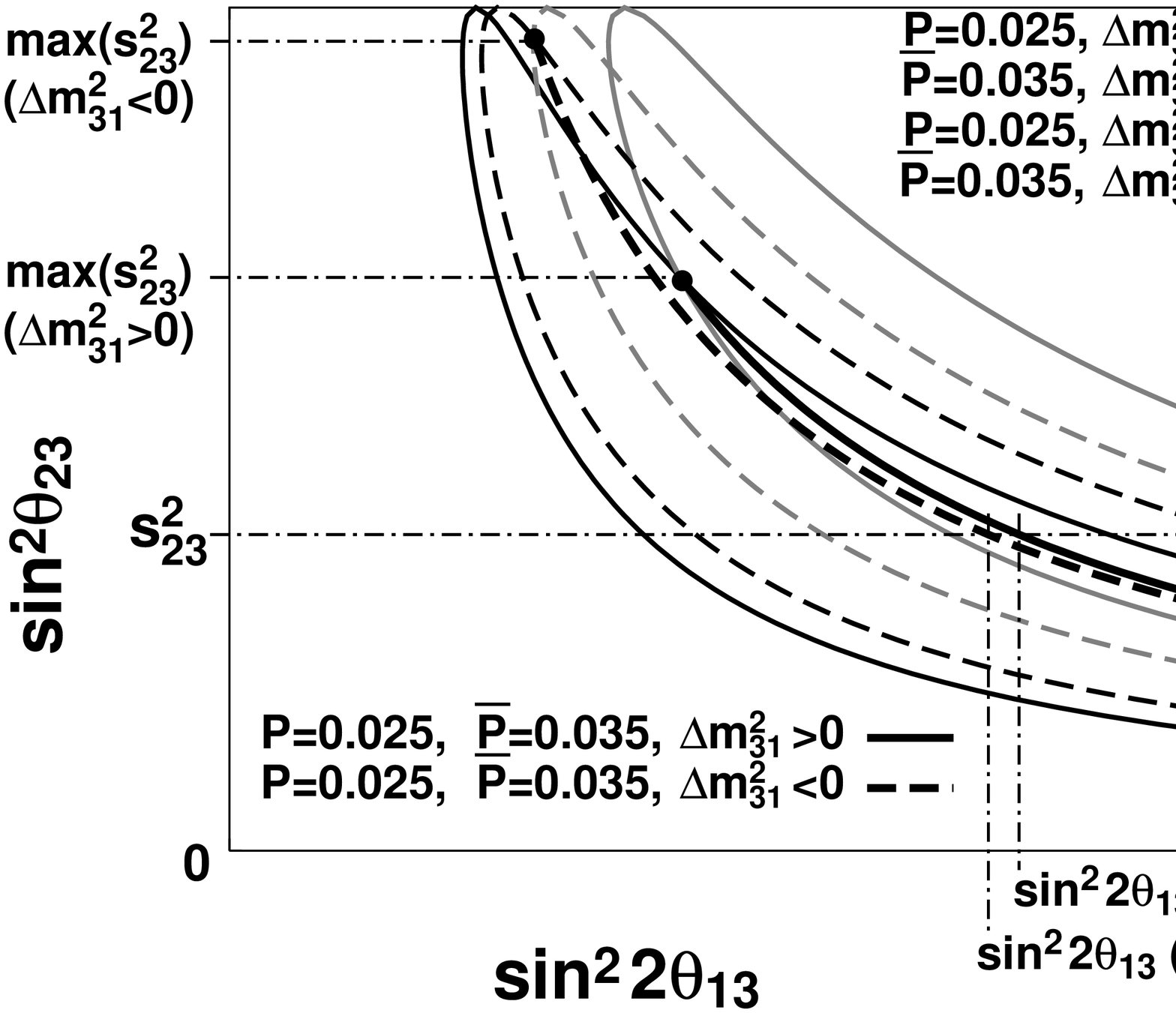}
\caption{{\scriptsize The contours which are given by
both $P$ and ${\bar P}$ for $\Delta m^2_{31}>0$
(the solid thick line)
and for $\Delta m^2_{31}<0$ (the dashed thick line) at
the oscillation maximum $|\Delta m^2_{31}L/4E|=\pi/2$.
These lie in the overlap of the region which is given by
$P$ only (the area which is bound by the thin black solid (dashed) line
for $\Delta m^2_{31}>0$ ($\Delta m^2_{31}<0$))
and the region which is given by ${\bar P}$ only
(the area which is bound by the thin gray solid (dashed) line
for $\Delta m^2_{31}>0$ ($\Delta m^2_{31}<0$)).
}}
\label{fig:om}
\end{figure}

\section{Parameter degeneracies}
It has been realized
that we cannot determine the oscillations parameters
$\theta_{jk}$, $\Delta m^2_{jk}$, $\delta$ uniquely
even if we know precisely the appearance probabilities
$P(\nu_\mu\rightarrow\nu_e)$ and
$P(\bar{\nu}_\mu\rightarrow\bar{\nu}_e)$ in a long
baseline accelerator experiment with an approximately
monoenergetic neutrino beam
due to so-called parameter degeneracies.
There are three kinds of the parameter degeneracies:
the intrinsic $(\theta_{13}, \delta)$
degeneracy, the degeneracy of
$\Delta m^2_{31}\leftrightarrow-\Delta m^2_{31}$, and
the degeneracy of $\theta_{23}\leftrightarrow\pi/2-\theta_{23}$.
Each degeneracy gives a two-fold solution, so
in total one has eight-fold solution if the
degeneracies are exact.  When these degeneracies
are lifted, there are eight solutions
for the oscillation parameters such as $\theta_{13}$ etc.,
and it will be important in future long baseline
experiments to discriminate the real solution
from fake ones.

To explain the parameter degeneracies,
let me consider the contours which are given
by $P\equiv P(\nu_\mu\rightarrow\nu_e)$
and ${\bar P}\equiv P(\bar{\nu}_\mu\rightarrow\bar{\nu}_e)$
at the same time in
the $\sin^22\theta_{13}$--$s^2_{23}$ plane.
If there were no matter effect $A\equiv \sqrt{2}G_FN_e=0$,
if the mass squared difference $\Delta m^2_{21}$ of the
solar neutrino oscillation were exactly zero, and if
$\theta_{23}$ is exactly $\pi/4$, then we would have
one solution with 8-fold degeneracy as is shown in
Fig.\ref{fig:degeneracy}(a).  If we lift the conditions
in the order
$(A=0, \Delta m^2_{21}=0,\theta_{23}-\pi/4=0)$
$\rightarrow$
$(A=0, \Delta m^2_{21}=0,\theta_{23}-\pi/4\ne0)$
$\rightarrow$
$(A=0, \Delta m^2_{21}\ne0,\theta_{23}-\pi/4\ne0)$
$\rightarrow$
$(A\ne0, \Delta m^2_{21}\ne0,\theta_{23}-\pi/4\ne0)$,
then the exact degeneracies are lifted as is depicted
in Figs.\ref{fig:degeneracy}(a) (one solution
with 8-fold degeneracy)
$\rightarrow$ 
(b) (two solutions
with 4-fold degeneracy) $\rightarrow$ (c) (four solutions
with 2-fold degeneracy) $\rightarrow$ (d) (eight solutions
without any degeneracy).
Furthermore, if we assume the neutrino energy to be
approximately monoenergetic and to satisfy the
oscillation maximum condition
$|\Delta m^2_{31}L/4E|=\pi/2$ as is the case
at the JHF experiment, then the contours look like
Fig.\ref{fig:degeneracy}(e) and in this case only
ambiguity which causes a problem is the $\theta_{23}$
degeneracy, since the intrinsic $(\theta_{13}, \delta)$
degeneracy is exact and there is little ambiguity due to
the $\mbox{\rm sgn}(\Delta m^2_{31})$
degeneracy because $|AL/2|\ll 1$.

Here let me explain briefly the behaviors of the contours
in Fig.\ref{fig:degeneracy}(e), which is shown in
Fig.\ref{fig:om} in detail.  Using the analytic approximate
formulae\cite{Cervera:2000kp}
in the case of the oscillation maximum,
\begin{eqnarray}
P
=x^2 f^2 - 2 x y f g
\sin\delta
+ y^2 g^2\,,\nonumber\\
\bar{P} =
x^2 \bar f^2 + 2 x y \bar f g \sin\delta
+ y^2 g^2,
\label{degene}
\end{eqnarray}
where $x\equiv s_{23}\sin2\theta_{13}$,
($f,\bar{f})\equiv {\cos(AL/2)/(1\mp AL/\pi) }$,
$g \equiv {\sin(AL/2) /  (AL/\pi)}$,
$x \equiv s_{23} \sin 2\theta_{13}$,
$y \equiv \epsilon\,c_{23} \sin 2\theta_{12}$ and
$\epsilon\equiv|\Delta m^2_{21}/\Delta m^2_{31}|$,
one can show for any given value of $s^2_{23}$
\begin{eqnarray}
\left.\sin^22\theta_{13}\right|_{\Delta m^2_{31}>0}
-\left.\sin^22\theta_{13}\right|_{\Delta m^2_{31}<0}
&=& {1 \over s^2_{23}}{1 \over f{\bar f}}
{f-{\bar f} \over f+{\bar f}}(P-{\bar P})\nonumber\\
&\simeq&{AL \over \pi s^2_{23}}(P-{\bar P}),
\label{eqn:diffsin213}
\end{eqnarray}
if $|AL/2|\ll 1$.
Therefore the thick solid and thick dashed lines are close
to each other in Fig.\ref{fig:om}
because the matter effect and
$\epsilon$ are both small.
Notice that $|P-{\bar P}|$ could be large
if $\epsilon$ were large even if $|AL/2|\ll 1$.
One can also show
\begin{eqnarray}
\left.\max\,s^2_{23}\right|_{\Delta m^2_{31}<0}
-\left.\max\,s^2_{23}\right|_{\Delta m^2_{31}>0}
&=& {1 \over \epsilon^2}{1 \over \sin^22\theta_{12}}
{1 \over g^2}{f-{\bar f} \over f+{\bar f}}(P-{\bar P})
\nonumber\\
&\simeq&{1 \over \epsilon^2}
{(2/\pi)^2 \over\sin^22\theta_{12}}
{AL \over \pi}(P-{\bar P})
\label{eqn:diffmaxs23}
\end{eqnarray}
if $|AL/2|\ll 1$.
(\ref{eqn:diffmaxs23}) indicates
that $\max\,s^2_{23}|_{\Delta m^2_{31}<0}
-\max\,s^2_{23}|_{\Delta m^2_{31}>0}$ is large
in Fig.\ref{fig:om} because a small quantity
$AL(P-{\bar P})$ is enhanced by a large factor
$1/\epsilon^2$.
In the case of the JHF experiment,
which has $L$=295km, $A\simeq$1/(1900km),
$\epsilon\simeq 1/35$, if $P$=0.025 and ${\bar P}$=0.035, then
the right hand side of (\ref{eqn:diffsin213}) $\simeq5\times10^{-4}$,
and the right hand side of (\ref{eqn:diffmaxs23}) $\simeq 0.3$.

\section{Reactor measurements of $\theta_{13}$}
In the three flavor framework
the disappearance probability of the reactor
neutrinos with a baseline less than 10km is given by

\begin{eqnarray}
P(\bar{\nu}_{e} &\rightarrow& \bar{\nu}_{e}) =1-
\sin^22\theta_{13}
\sin^2\left({\Delta m^2_{31}L \over 4E}\right),
\label{Pvac}
\end{eqnarray}
to a good approximation.  Hence
reactor measurements are free from the
ambiguity due to the matter effect, the CP phase $\delta$
and $\theta_{23}$.  It has been
shown\cite{Minakata:2002jv,suekane}
that a reactor measurement at the Kashiwazaki-Kariwa
nuclear power plant potentially has sensitivity down to
$\sin^2{2 \theta_{13}}\simeq 0.02$.

\begin{figure}
\includegraphics[scale=0.3]{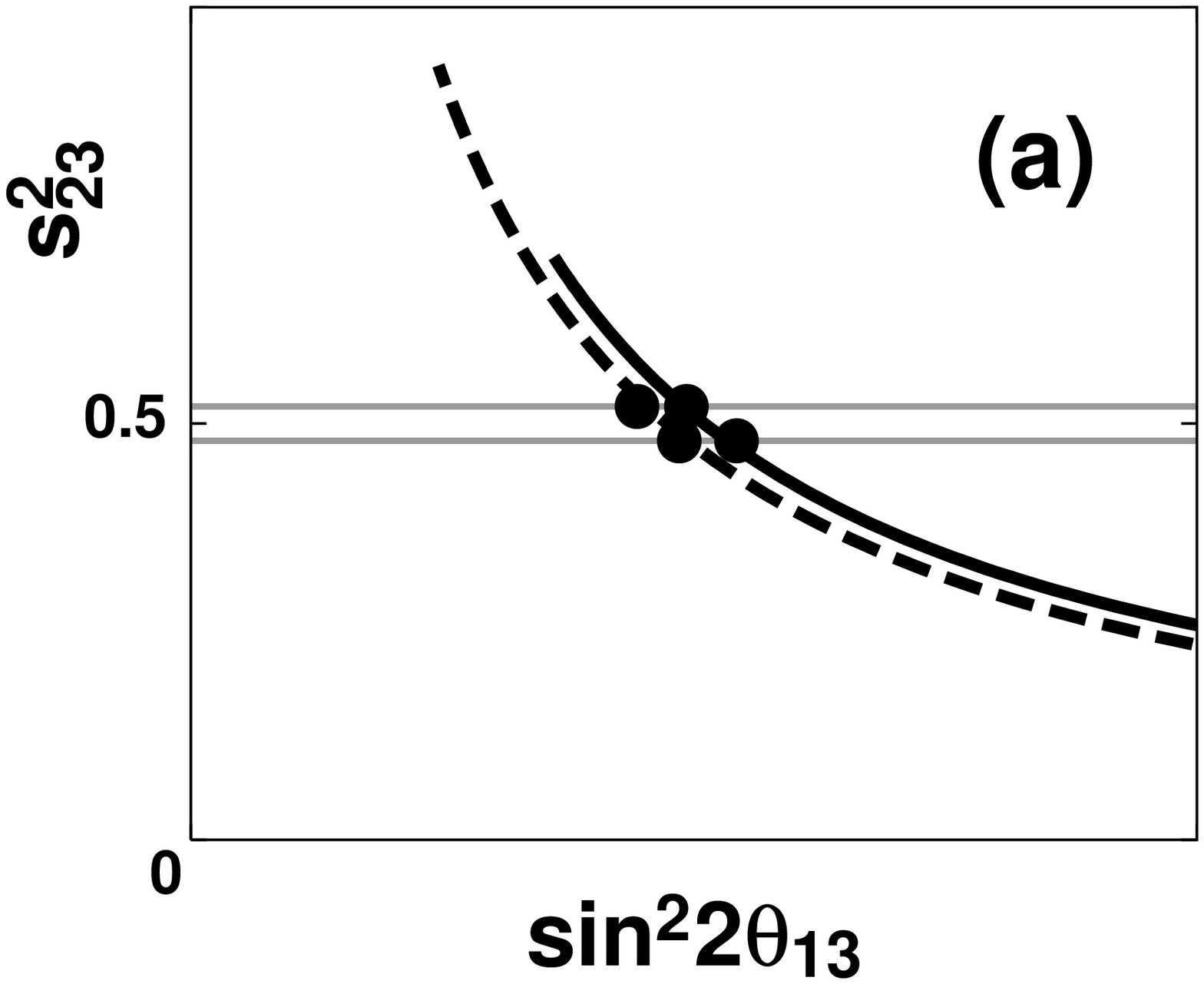}
\vglue -4.3cm
\hglue 6.cm
\includegraphics[scale=0.3]{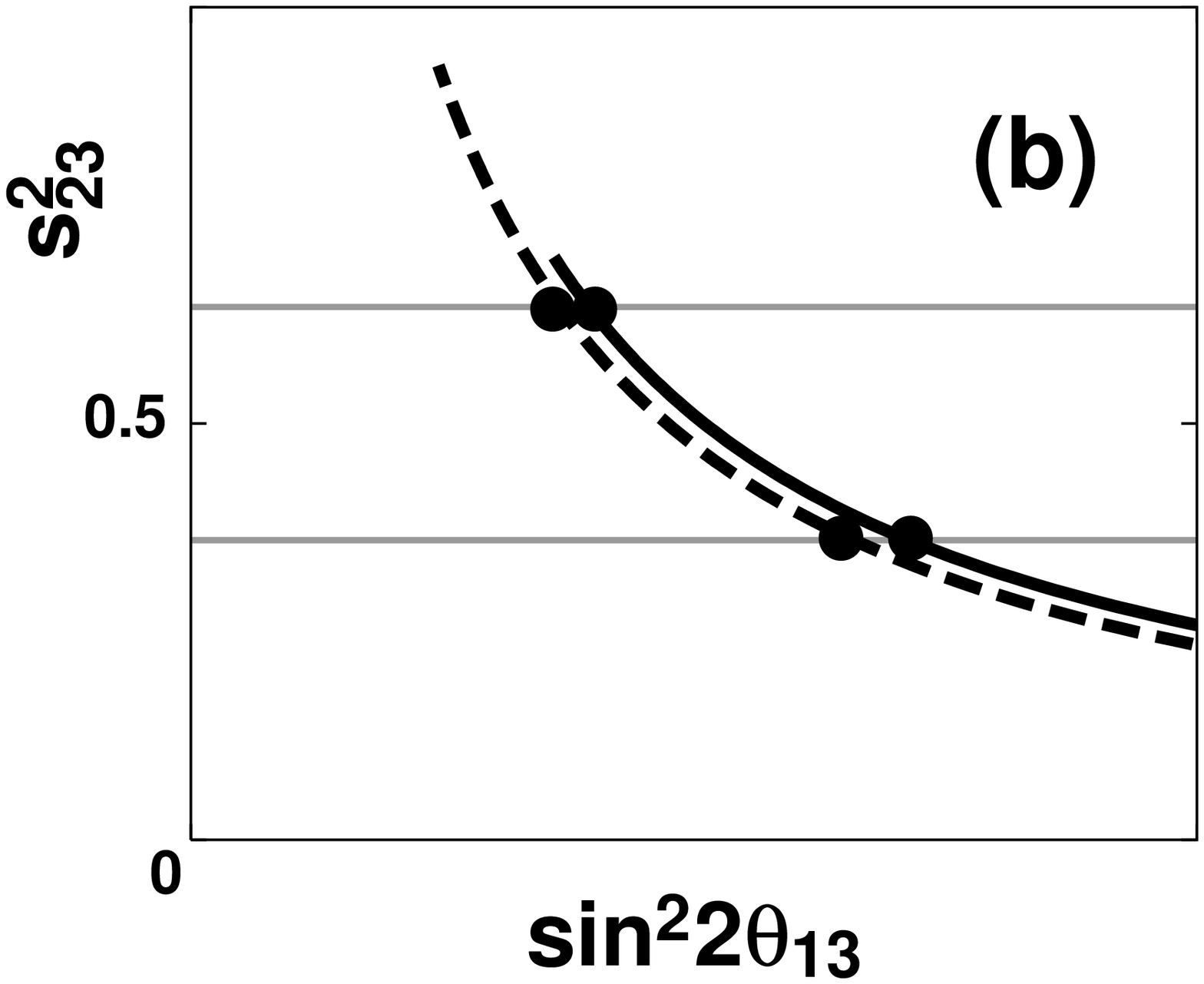}
\vglue .5cm
\caption{{\scriptsize Situations of the long baseline
accelerator experiment at the oscillation maximum
($|\Delta m^2_{31}L/4E|=\pi/2$):
(a) the case with $\theta_{23}\simeq \pi/4$,
(b) the case with $\theta_{23}\ne \pi/4$.}}
\label{fig:jhfre}
\end{figure}

\begin{figure}
\vglue -5.5cm
\hglue 1.5cm
\includegraphics[scale=0.45]{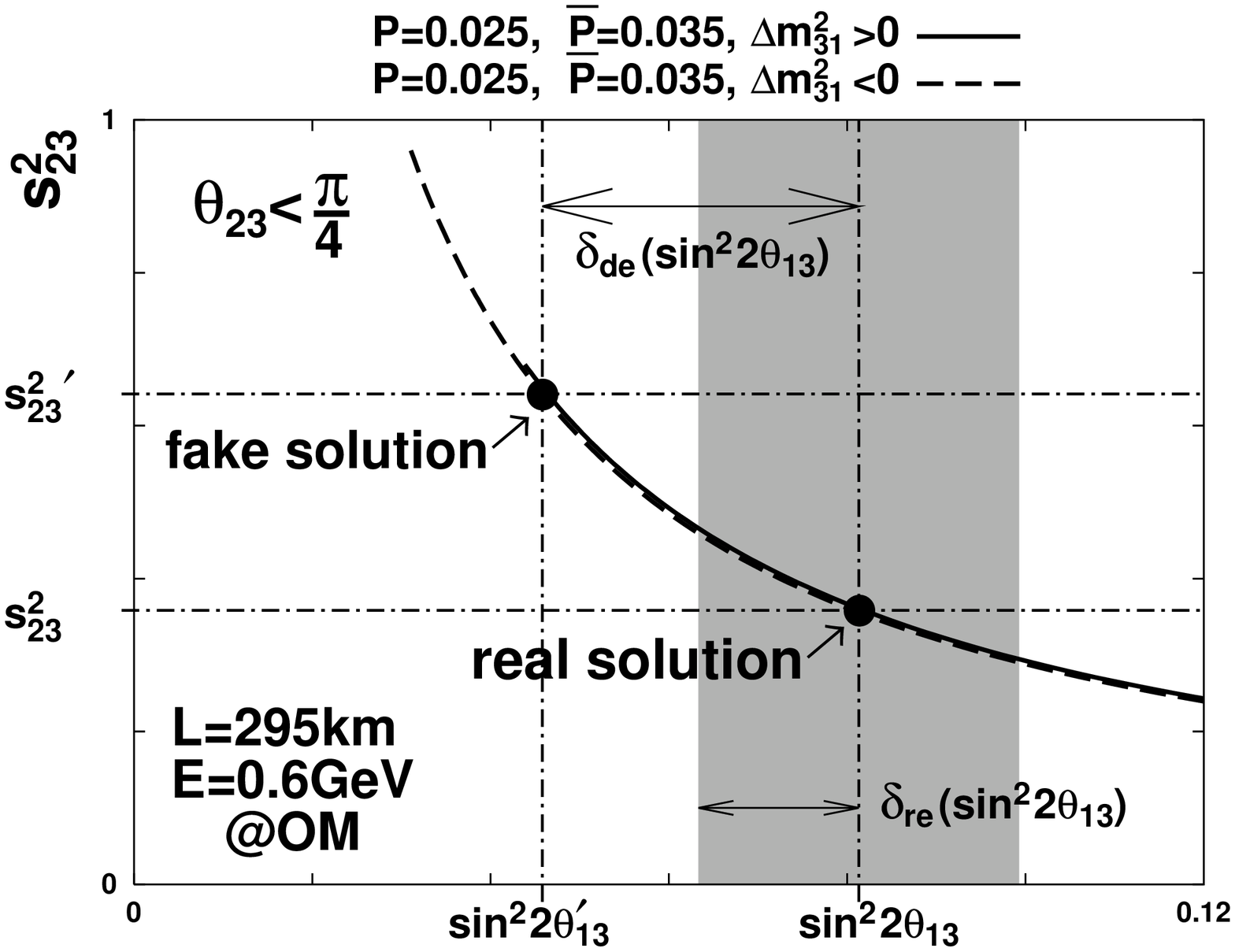}
\vglue -4.cm
\hglue 1.5cm
\includegraphics[scale=0.45]{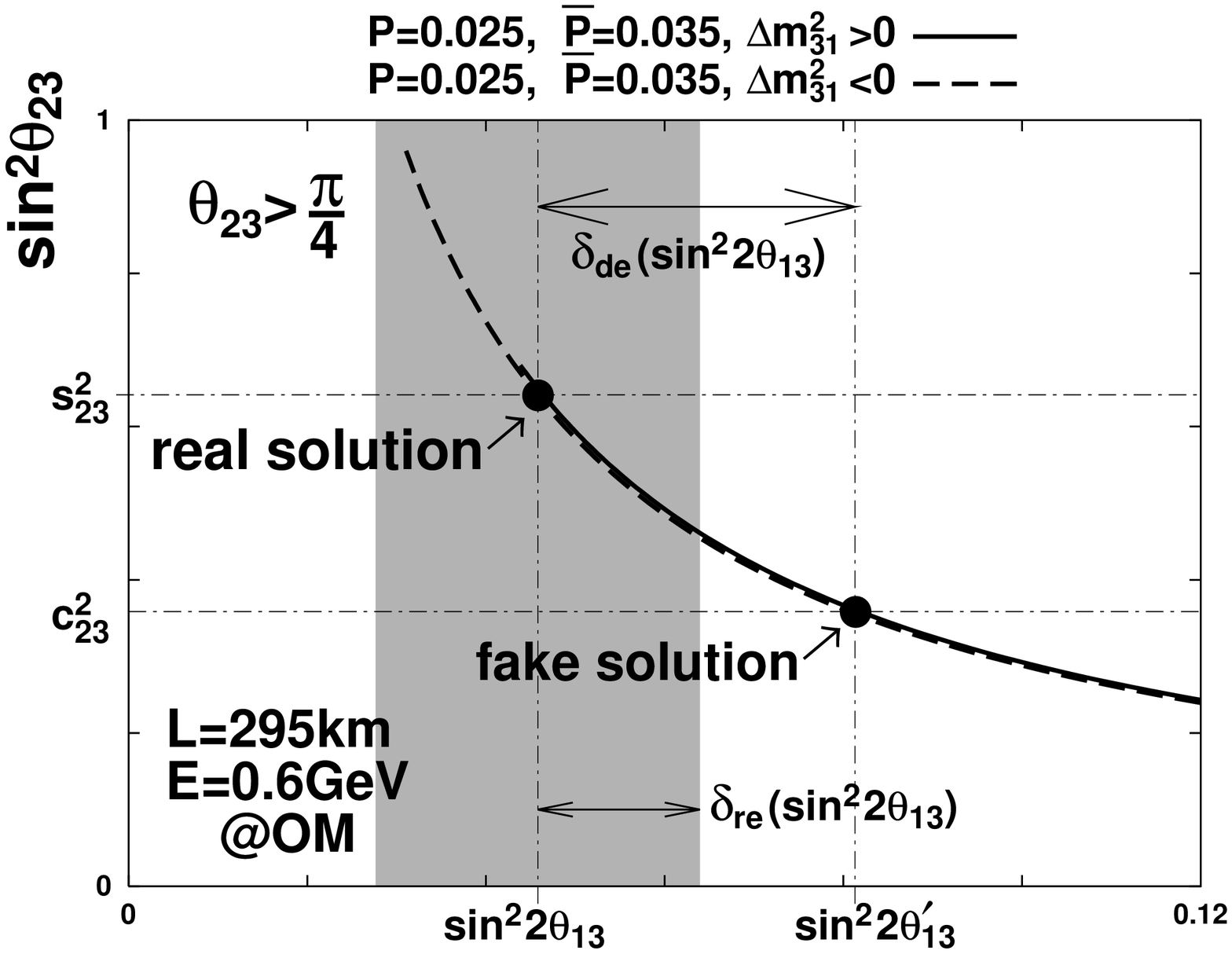}
\vglue 1.5cm
\caption{{\scriptsize Resolution of the $\theta_{23}$
degeneracy by combining the results 
of the reactor and the long baseline accelerator experiments
at the oscillation maximum:
(a) the case with $\theta_{23}< \pi/4$,
(b) the case with $\theta_{23}> \pi/4$.
The shaded region stands for the error in $\sin^22\theta_{13}$
of the reactor experiment.}}
\label{fig:fig5}
\end{figure}

\begin{figure}
\vglue .7cm
\hglue -1.0cm
\includegraphics[scale=0.45]{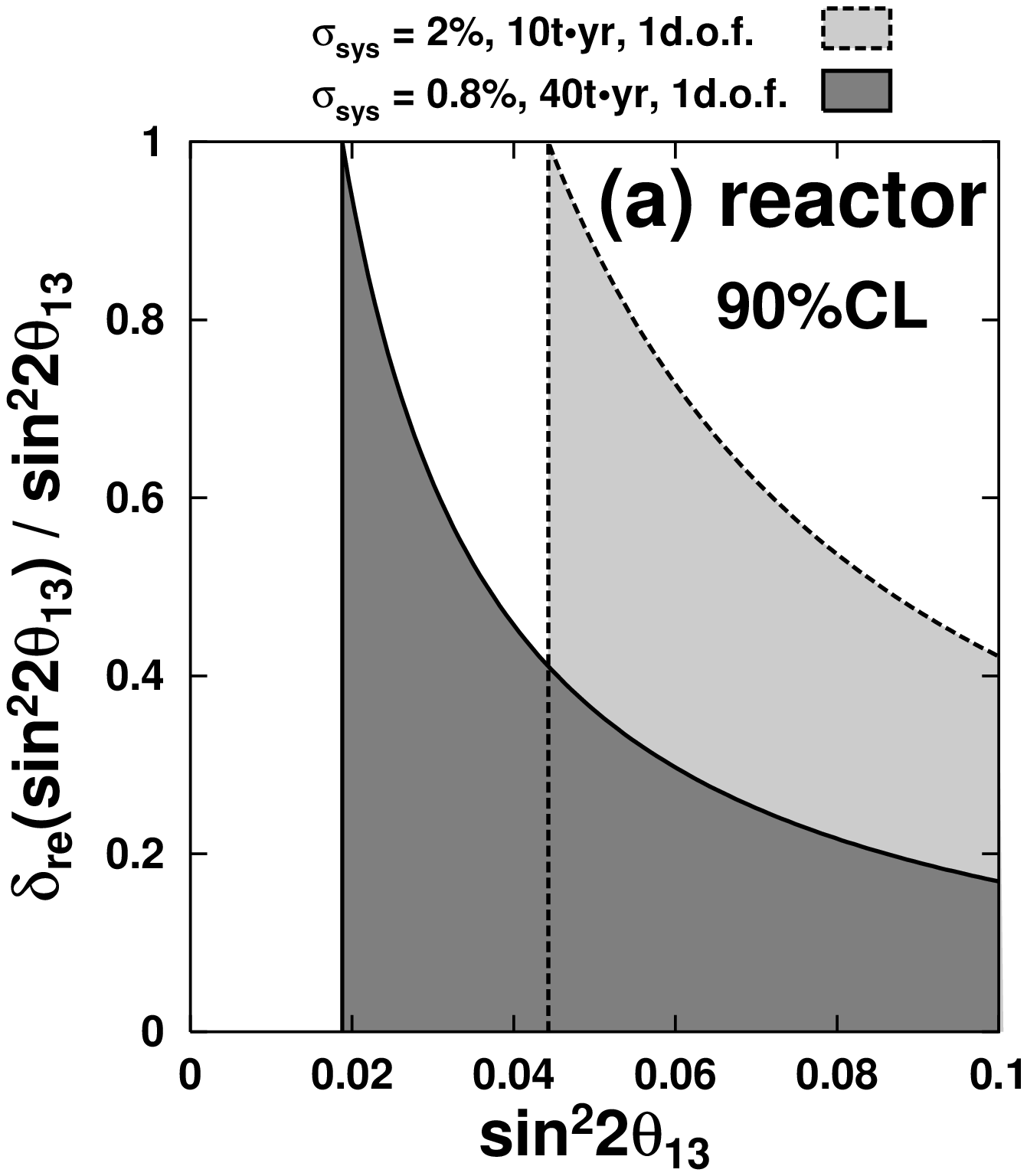}
\vglue -6.4cm
\hglue 5.5cm
\includegraphics[scale=0.45]{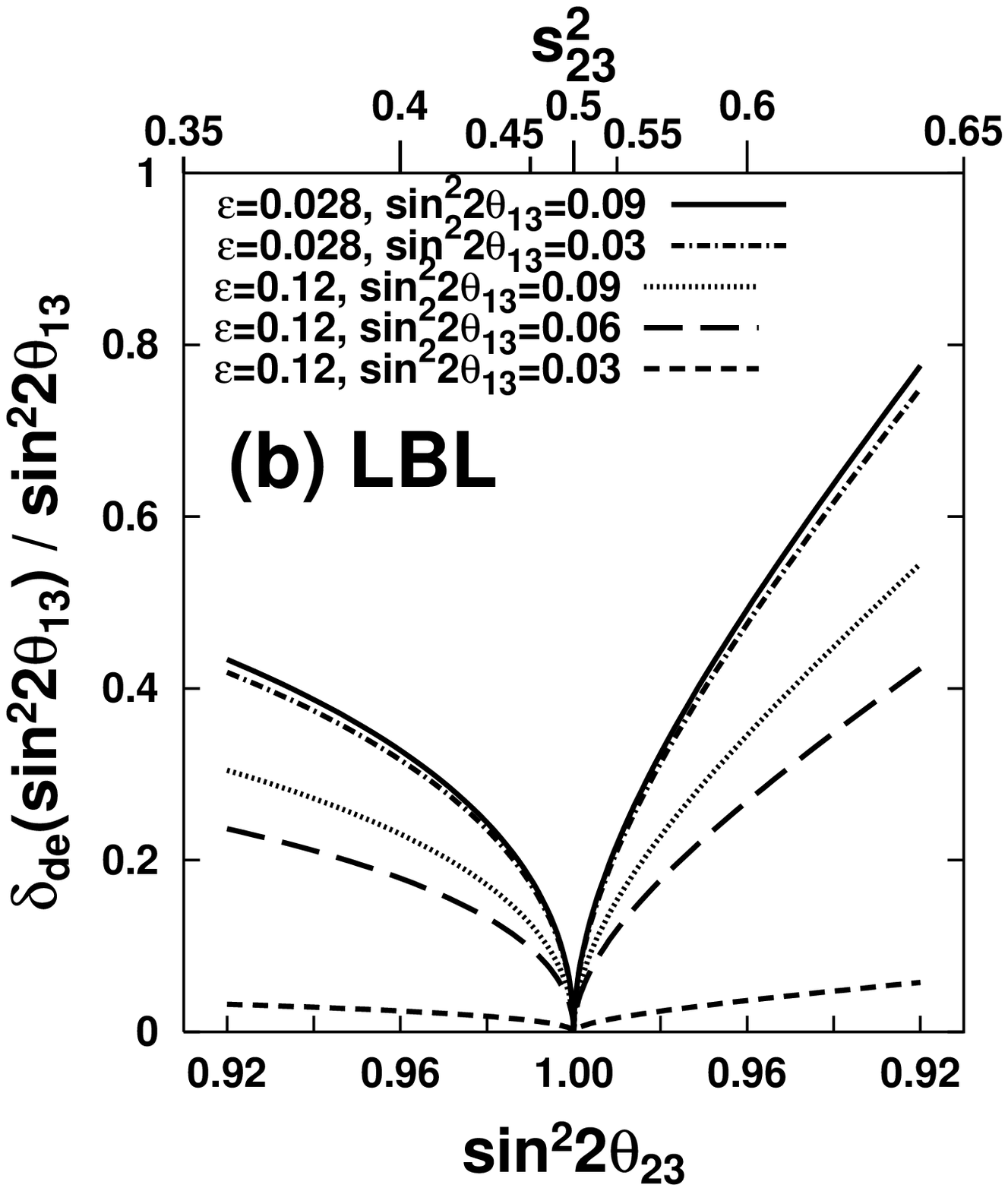}
\vglue .1cm
\caption{{\scriptsize (a) Errors in the reactor experiment.
(b) Ambiguities due to the $\theta_{23}$ degeneracy
in the long baseline accelerator experiment at the
oscillation maximum, where
the horizontal axis is defined in such a way that
it is near in $\sin^22\theta_{23}$.
$\epsilon\equiv|\Delta m^2_{21}/\Delta m^2_{31}|$=0.028
is the best fit case for the solar and atmospheric
neutrino data where
$\delta_{de}(\sin^22\theta_{13})/\sin^22\theta_{13}$
is approximated by $|1-\tan^2\theta_{23}|$ for almost
all the values of $\sin^22\theta_{13}$,
while $\epsilon=0.12$ is the most
pessimistic case within the 90\%CL allowed region
for the solar and atmospheric neutrino data,
for which approximation by $|1-\tan^2\theta_{23}|$
is not good.
}}
\label{fig:fig6}
\end{figure}

\section{Resolution of the $\theta_{23}$ ambiguity
by the LBL and reactor experiments}
Once we have the results from, say, the phase 2 of
the JHF experiment on $P(\nu_\mu\rightarrow\nu_e)$
and $P({\bar \nu}_\mu\rightarrow{\bar \nu}_e)$ and
the reactor measurement of
$P({\bar \nu}_e\rightarrow{\bar \nu}_e)$,
there are two possibilities.  One is the case
where $\sin^22\theta_{23}$ is close to 1, as is shown in
Fig.\ref{fig:jhfre}(a).  In this case one could not
resolve the ambiguity within the experimental error
nor would one have to worry about the ambiguity,
as the difference is small.
On the other hand, if $\sin^22\theta_{23}$ turns out
to be away from 1 as is depicted in Fig.\ref{fig:jhfre}(b),
then the reactor result may enable us to resolve
the $\theta_{23}$ ambiguity.

As is shown in Fig.\ref{fig:fig5}(a) and (b), if the ambiguity
$\delta_{de}(\sin^22\theta_{13})\equiv|\sin^22\theta_{13}^{\prime}
-\sin^22\theta_{13}|$ due to the degeneracy,
where $\theta_{13}$ and
$\theta_{13}^{\prime}$ stand for the values of
$\theta_{13}$ for the real and fake solutions,
is larger than the error
$\delta_{re}(\sin^22\theta_{13})$ of the reactor
measurement, then we can resolve the $\theta_{23}$ ambiguity.
$\delta_{de}(\sin^22\theta_{13})$
can be computed from the equation (26) in
Ref.\cite{Barger:2001yr}, and we obtain
\begin{eqnarray}
&{\ }&
{\delta_{de}(\sin^22\theta_{13}) \over
\sin^22\theta_{13}} 
\nonumber\\
&=&
\left|1-\tan^2\theta_{23}\right|
\left[1 + {\epsilon^2 \over \sin^22\theta_{13}}
{\tan^2\left(aL/2\right) \over \left(aL/\pi\right)^2}
[1-({aL \over\pi})^2 ]
\sin^22\theta_{12}\right].
\label{eqn:th23}
\end{eqnarray}
Hence $\delta_{de}(\sin^22\theta_{13})/\sin^22\theta_{13}$
can be approximated by $|1-\tan^2\theta_{23}|$ unless
$\epsilon$ is large such as 0.1.  As for the error
of the reactor measurements, we have
$\delta_{re}(\sin^22\theta_{13})\simeq 0.018$ for
any value of $\sin^22\theta_{13}$ in the case
with the systematic error 0.8\% and 40t$\cdot$yr.
$\delta_{re}(\sin^22\theta_{13})/\sin^22\theta_{13}$ and
$\delta_{de}(\sin^22\theta_{13})/\sin^22\theta_{13}$ are
shown in Fig.\ref{fig:fig6}(a) and (b), and from
these figures one can read off
the region where the $\theta_{23}$ ambiguity is resolved.
In general the $\theta_{23}$ ambiguity is resolved
if $\sin^22\theta_{13}$ and $1-\sin^22\theta_{23}$ are
both large.

\section{Summary}
In this talk I have shown that the 8-fold parameter
degeneracy in long baseline experiments can be
visualized in the $\sin^22\theta_{13}$--$s^2_{23}$ plane.
I have demonstrated that one may be able to resolve
the $\theta_{23}$ ambiguity by combining the results
of the JHF experiment at the oscillation maximum and
a reactor experiment whose sensitivity for $\sin^22\theta_{13}$
is 0.02, if $\sin^22\theta_{13}$ and $\cos^22\theta_{23}$ are
both relatively large.  This scenario offers one of the
strategies which
are expected to resolve the ambiguities due to the
parameter degeneracies.\footnote{
As other alternatives to resolve the ambiguities,
Parke\cite{parke} proposed to measure both
$P(\nu_\mu\rightarrow\nu_e)$ and
$P({\bar \nu}_\mu\rightarrow{\bar \nu}_e)$ at JHF
off the oscillation maximum and at NuMI at
the oscillation maximum, whereas
Donini\cite{donini} discussed the measurements of
the golden channel $P(\nu_e\rightarrow\nu_\mu)$
and the silver channel $P(\nu_e\rightarrow\nu_\tau)$
at a neutrino factory.}


\begin{thebibliography}{0}
\bibitem{Minakata:2002jv}
H.~Minakata, H.~Sugiyama, O.~Yasuda, K.~Inoue and F.~Suekane,
arXiv:hep-ph/0211111.

\bibitem{suekane}
F.~Suekane, 
\fussy
talk at {\em The 4th Workshop on "Neutrino Oscillations
and Their Origin"}, February 10-14, 2003, Kanazawa, Japan\\
(http://www-sk.icrr.u-tokyo.ac.jp/noon2003/transparencies/11/Suekane.pdf).

\bibitem{Cervera:2000kp}
A.~Cervera, A.~Donini, M.~B.~Gavela, J.~J.~Gomez Cadenas, P.~Hernandez, O.~Mena and S.~Rigolin,
Nucl.\ Phys.\ B {\bf 579}, 17 (2000)
[Erratum-ibid.\ B {\bf 593}, 731 (2001)]
[arXiv:hep-ph/0002108].

\bibitem{Barger:2001yr}
V.~Barger, D.~Marfatia and K.~Whisnant,
Phys.\ Rev.\ D {\bf 65}, 073023 (2002)
[arXiv:hep-ph/0112119];

\bibitem{parke}
S.~Parke, 
\fussy
talk at {\em The 4th Workshop on "Neutrino Oscillations
and Their Origin"}, February 10-14, 2003, Kanazawa, Japan\\
(http://www-sk.icrr.u-tokyo.ac.jp/noon2003/transparencies/11/Parke.pdf).

\bibitem{donini}
A.~Donini, 
\fussy
talk at {\em The 4th Workshop on "Neutrino Oscillations
and Their Origin"}, February 10-14, 2003, Kanazawa, Japan\\
(http://www-sk.icrr.u-tokyo.ac.jp/noon2003/transparencies/11/Donini.pdf).

\end{thebibliography}
\end{document}